\documentclass[prl, twocolumn, showpacs, preprintnumbers, amsmath, amssymb, superscriptaddress, aps]{revtex4}
\usepackage{graphicx}
\usepackage{dcolumn}
\usepackage{color}
\usepackage{bm}
\def\he4{$^4$He}
\def\hee3{$^3$He}
\def\Am3{\AA$^{-3}$}
\def\beq{\begin{equation}}
\def\eeq{\end{equation}}
\begin{document}

\author{D. Aleinikava}

\author{E. Dedits}

\author{A.B. Kuklov}
\affiliation{Department of Engineering Science and Physics,
CSI, CUNY, Staten Island, NY 10314, USA}

\author{D. Schmeltzer}
\affiliation{Department of Physics, CCNY
CUNY, New Yourk, NY 10031, USA}


\title{Mechanical and superfluid properties of dislocations in solid \he4 }

\date{\today}

\begin{abstract}
Dislocations are shown to be smooth at zero temperature because of the effective Coulomb-type interaction between kinks. 
Crossover to finite temperature rougnehing is suggested to be a mechanism responsible for the softening of \he4 shear modulus recently observed by Day and Beamish (Nature, {\bf 450}, 853 (2007)). 
We discuss also that strong suppresion of superfuidity along the dislocation core by thermal kinks can lead to locking in of the mechanical and superfluid responses.   
\end{abstract}

\pacs{67.80.bd, 67.80.-s, 05.30.Jp, 61.72.Ff}

\maketitle

Network of superfluid (SF) dislocations \cite{Schevchenko} is a most likely scenario for the supersolid  response of solid \he4 on rotation \cite{KC} at temperatures $T\leq 0.2K$, at least, for values of the SF fraction $\rho_s(T)\leq 0.01$. The proof \cite{noSFS} that ideal crystal of \he4 cannot be a supersolid \cite{Andreev69} and the observations of  SF cores of some dislocations in first principle simulations \cite{screw} 
  put the model \cite{Schevchenko} on a solid ground. Dislocations are building blocks of most topological defects. Thus, understanding their properties is of primary importance.

Classical and quantum mechanical behavior of dislocations was addressed by many investigators in the past \cite{old_disl}. Quantum roughening of dislocations has been proposed to be  important for inducing supersolidity \cite{deGennes}. Remarkable resemblance between shear modulus temperature dependence $G(T)$ and $\rho_s(T)$ was uncovered recently \cite{Beamish}.  

Here we study quantum behavior of a single dislocation and its crossover to classical regime. We also discuss a possibility of suppression of SF along the core \cite{screw}  
 by geometrical kinks on dislocation line. Our findings are that such simple model, which ignores collective effects of the dislocation network, can naturally explain the key features of the experiment \cite{Beamish}.  Some of our preliminary results have been presented in Ref.\cite{Trieste}.

{\it Model}. Edge dislocation moving along its gliding plane $(x,y)$ is modeled as a string characterized by some displacement field $y(x,t)$ with kinetic and tension energies $H_s =\int dx[(n_1/2) (\nabla_t y)^2 + (V^2_d n_1/2)(\nabla_x y)^2]$ and subjected to Peierls potential $U_P(y)= \int dx\, u\cos(2\pi y(x,t)/b)$ with some amplitude $u$ (see in Ref.\cite{Kosevich}). Here $n_1$ is 1d mass density; $V_d$ -- speed of sound along the string;  $b$ denotes Burgers vector. The Hamiltonian $H_s$ alone was successfully used for describing dynamical properties of many materials at high temperatures \cite{Granato}. At low $T$, however, the Peierls potential is essential. Furthermore, static kinks interact with each other through effective Coulomb potential $\tilde{V}_C(x)$ \cite{Hirth,Kosevich} (induced by exchanging bulk sound phonons with bulk velocity $V_b \approx V_d$). There is also  long-range contribution to the effective mass \cite{Kosevich}. So, we modify $H_s+U_P$ by including the long-range interaction with retardation effects and obtain $\tilde{V}_C$ from the bulk phonon propagator 
integrated over the directions of momenta perpendicular to the core.
Thus, full action $S$ in imaginary time is given as 
\begin{eqnarray}
\frac{S}{\hbar}&=&  \sum_{\omega, q_x} \left[ \frac{1+ V_C(q_x,\omega)}{2K}(\omega^2 +q_x^2)|\theta_{\omega,q_x}|^2 \right] 
\label{S2} \\
&-& \int_0^{N_x} dx \int_0^\beta d t\, u \cos(\sqrt{16\pi}\theta),
\label{UP} \\
& &V_C(q_x,\omega)=C\ln\left(1+ \frac{q_0^2}{\nu\omega^2 +  q_x^2}\right)
\label{VC}
\end{eqnarray}   
in units of typical cell size $x_0\approx 4\AA$ of \he4. Unit of time is $\tau_0=x_0/V_d$ so that $\beta =T_0 /T, \, T_0=\hbar V_d/x_0 \approx 5K$ for $V_d\approx 200$m/s; and expansion is performed over Matsubara frequencies $\omega = (2\pi/\beta) n, \, n=0,1,2,...$ and spatial wavevector $q_x$ along the core of length $L_x=N_x x_0$ ;   $q_0\approx 1$ ; $K= \pi \hbar/(4 n_1 V_d b^2)$; we have chosen the rescaling $y=2b \theta/\sqrt{\pi}$, with $\theta_{\omega,q_x}$ standing for the Fourier component of $\theta$; $\nu= (V_d/V_b)^2$ (we will be using $\nu=1$). $C$ describes relative energy of the bulk deformations with respect to the core. The first one is usually the largest \cite{Hirth,Kosevich}. Thus, in real materials $C>1$. 
In calculating the partition function $Z=\int \, Dy \, \exp(-S/\hbar)$ we use periodic boundary conditions $\theta(x+L,t)=\theta(x,t),\, \theta(x,t+\beta)=\theta(x,t)$. 

It is important to note that, for solid \he4 , the Luttinger parameter $K$ in the term (\ref{S2}) is close to unity, that is, for $C=0$ the dislocation is close to Berezinskii-Kosterlitz-Thouless (BKT) transition (see in Ref.\cite{BKT}), so that it is in {\it quantum rough} state for $K>1/2$. Below we will show that arbitrary small $C$ inevitably brings the system into the {\it quantum smooth} state as $L_x\to \infty$ for any $K$.

Following Granato-L\"ucke \cite{Granato}, change of $G$ can be related to a response of a typical dislocation segment of the dislocation network. If the whole network is viewed as a set of blocks of free segments of sizes $L_x,L_y,L_z$ along the corresponding orthogonal axes, a displacement $y(x,t)$ under the force $f=\sigma_{zy} b$, where $\sigma_{zy}$ stands for the stress tensor \cite{Landau}, results in a strain of the block $u_{zy}\approx b\langle y\rangle/(L_yL_z)$. Thus, $\langle y\rangle\propto \sigma_{zy}$ and, accordingly, $u_{zy}\propto \sigma_{zy}$. This leads to
\begin{eqnarray}
\frac{1}{G(T,L_x)}=\frac{1}{G_{el}} + \frac{n_d}{\hbar} \int^\beta_0 dt \int_0^{L_x} dx \langle y(x,t) y(0,0)\rangle
\label{G} 
\end{eqnarray}
where $G_{el}$ stands for the elastic modulus of ideal crystal; 
$n_d= b^2/(L_yL_z)\approx b^2/L_x^2$ is the dislocation density (in units of $b$), provided $L_y\approx L_z \approx L_x$. At large $T$, where the Peierls potential is irrelevant, the system is Gaussian and the high-$T$ modulus $G_\infty \equiv G(\infty,L_x)$ obeys 
\begin{eqnarray}
\frac{1}{G_\infty}=\frac{1}{G_{el}} + \frac{n_d}{q^2_LV^2_dn_1(1+CV(q_L,0))},
\label{Gm} 
\end{eqnarray}
where $q_L=2\pi/L_x$ for chosen boundary conditions and, in general, $q_L\propto 1/L_x$, so that the correction to the modulus is only weakly dependent on $L_x$: $\propto 1/(1+2C\ln (L_x/b))$. Absence of quantum roughening (proved below) implies that the Peierls potential is always relevant at $T=0$ so that the linear response is dominated by the lowest term in the expansion of $\cos(...)$ in Eq.(\ref{UP}). Thus, for $G_0\equiv G(0,L_x)$ we find 
\begin{eqnarray}
\frac{1}{G_0}=\frac{1}{G_{el}} + \frac{n_db^2}{(2\pi)^2 u}\approx \frac{1}{G_{el}}
\label{G0} 
\end{eqnarray}
For $L_x/b >> \sqrt{G_{el}b^2/u}/(2\pi)$.
At finite $T$, the dislocation undergoes a crossover from quantum smooth, Eq.(\ref{G0}), to classically rough state, Eq.(\ref{Gm}). 

{\it Heuristic argument against quantum roughening}. 
The system (\ref{S2},\ref{UP},\ref{VC}) can be mapped on the 2D classical gas model characterized by the interaction (in Fourier) between the integer charges which are dual variables to kinks \cite{BKT}:
\begin{equation}
\tilde{U}_{CG}= \frac{16\pi K}{(\omega^2+q_x^2)(1+V_C(q_x,\omega))}
\label{UCG}
\end{equation}
 Given the low $\omega , q_x$ asymptotic of $V_C$ in Eq.(\ref{VC}), the space-time asymptotic can be found as
$U_{CG}(r)\sim \ln(1+2C\ln r)/C \sim \ln \ln r, \, \, r=\sqrt{t^2 + x^2}$. That is, {\it slower} than $\sim \ln r$. Since the entropic contribution scales as $\sim - \ln r$, the free energy of a pair of charges $U_{CG}(r) - \ln r \to -\infty$ becomes unbounded from below. According to the Kosterlitz-Thouless argument this implies deconfinement of pairs of charges for {\it arbitrary} small $C$, that is, the {\it smooth} state without zero-point kinks.

{\it RG argument}. We construct RG flow equations in the one-loop approximation for $u$, $K$ and $C$  by considering them as scale dependent. Then, introducing the rescaling variable $l$, we find 
\begin{eqnarray}
&  &\frac{du}{dl}= 2u\left(1-\frac{2K}{1+ 2Cl}\right),
\label{u}\\  
&  &\frac{dK}{dl}= - \frac{K^3u^2}{(1+ 2Cl)^2},
\label{K}\\
&  &\frac{d(C/K)}{dl}= 0.
\label{C}
\end{eqnarray}  
An elementary analysis shows that, regardless of the initial conditions, the flow is always toward $K(l) \to 0, u(l)\to \infty$, that is, to the gapped state. This conclusion is similar to the result \cite{Girvin} for the case of dipole-dipole interactions in Ising model which also maps on Sine-Gordon model. We note, however, RG equations (\ref{u},\ref{K},\ref{C}) have only qualitative value and, as a comparison with the Monte Carlo (MC) simulations show, do not describe quantitatively the actual flow. 

{\it MC simulations}. 
Treating the term (\ref{UP}) in the Villain approximation (see in Ref.\cite{BKT}), the action (\ref{S2},\ref{UP},\ref{VC}) is reformulated as
\begin{eqnarray}
S=  \frac{1}{2}\sum_{i,j,}  U_d(\vec{x}_i-\vec{x}_j) \vec{J}_i\vec{J}_j 
\label{S_J} 
\end{eqnarray}   
in terms of the conserving integer currents $\vec{J}_i\equiv J^{(\nu)}_i$, $\nu=\pm \hat{x},\pm \hat{t}$, defined on each bond coming out of a given site $i$ of the space-time lattice, with $\pm \hat{x}, \pm \hat{t}$ being respective unit vectors. 
The Fourier of $U_d(\vec{x})$ is 
\begin{eqnarray}
\tilde{U}_d(q_x,\omega)= \frac{ (2\pi)^2}{(\omega^2+q_x^2)(\alpha^{-1} + \tilde{U}_{CG}(q_x,\omega))}. 
\label{U_V} 
\end{eqnarray}   
with $\alpha=u \Delta_x \Delta_t$ and $\Delta_x, \, \Delta_t$ being respective space-time sizes of the discretization cell (in units $x_0,\tau_0$).
Eq.(\ref{G}) in the $J$- variables becomes:
\begin{eqnarray}
G(T,L_x)=\frac{G_0}{1 +  \gamma \kappa(T,L_x)},\,\, \kappa\equiv \frac{\langle W^2_t\rangle}{TL_x}  
\label{Gdual} 
\end{eqnarray}
in the limit $L_x\to \infty$. Here $\gamma = b^2 n_dL_x^2 G_0 =const$; 
$ W_t=\sum_i J^{(\hat{t})}_i/N_t,\,\, N_t=\beta/\Delta_t$ is the total current in the $t$-direction, so that $\kappa$ is the compressibility of the  $J$-current model (\ref{S_J}).

\begin{figure}
\centerline{\includegraphics[angle = 0,
width=1.2\columnwidth]{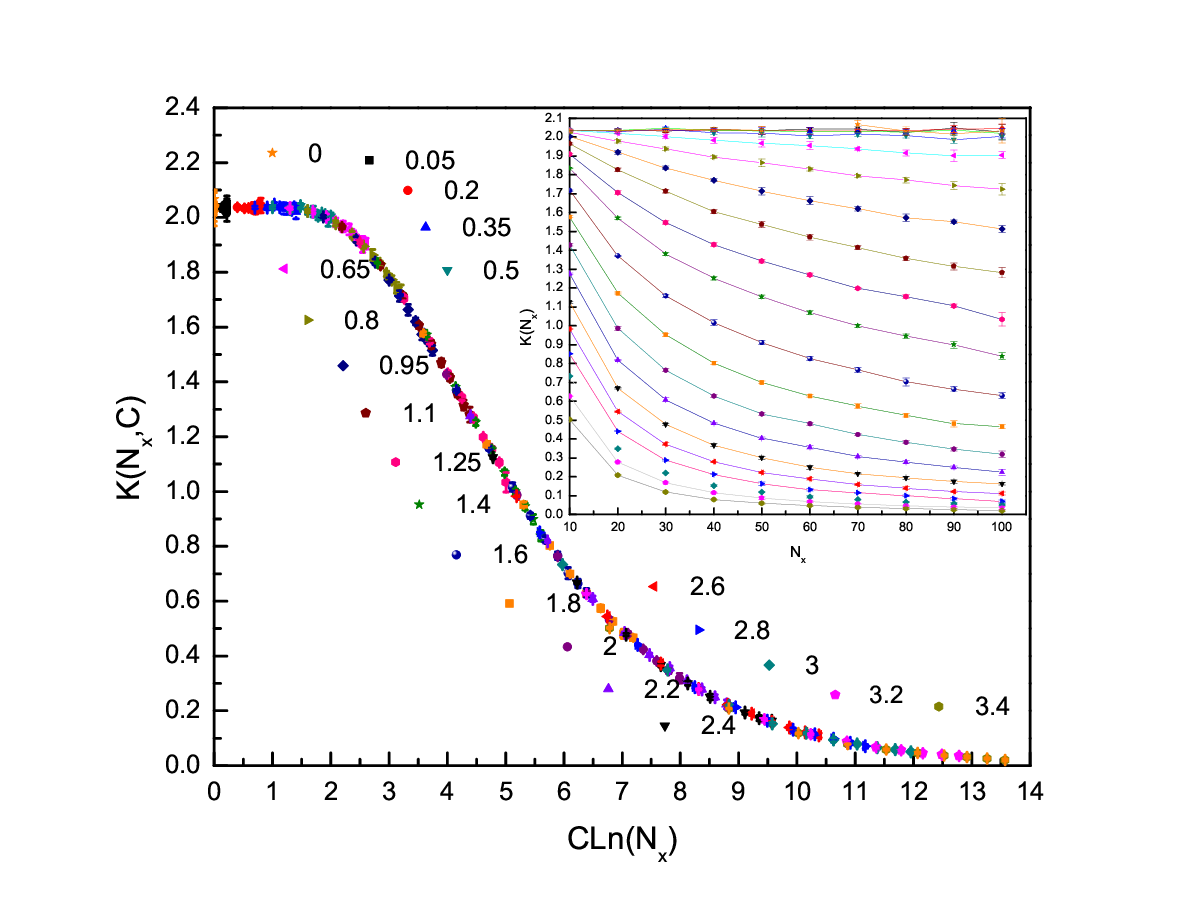}} \caption{(Color online)
Stiffness $K(N_x,C)$ at $T=0$ as a function of $\xi=C\ln N_x$. Shown numbers are values of $C$.  Chosen parameters are $K=1.6$, $\alpha =0.01$. 
Inset: family of curves $K(N_x,C)$ for various $C$.} \label{coll}
\end{figure}

MC simulations of the action (\ref{S_J}) have been conducted by the Worm Algorithm \cite{WA}.
We used isotropic lattice $N_t=N_x$ with $ \Delta_x=1$ and $\Delta_t=1$
, in progression of sizes and measured $\kappa$ as a function of $N_x, \, C$ for $K=1.6$ (well above the BKT critical point $K=0.5$ for $C=0$). The renormalized stiffness at $q_x=2\pi/N_x$ for $N_x >> 1$ is defined as $K(N_x,C)= \kappa \cdot (1+V_C(q_x,0))$. For given parameters and various values of $C$ all the data (shown in the inset of Fig.\ref{coll}) can be collapsed on a single master curve $K(N_x,C)= \tilde{K}(\xi)$ where $\xi=C\ln N_x $ and $\tilde{K}(\xi)=K(\xi,1)$, Fig.\ref{coll}, which asymptotically reaches zero.  This concludes our proof that a single dislocation is in the {\it quantum smooth} state for an arbitrary value of the long-range interactions $C$. 
 
\begin{figure}
\centerline{\includegraphics[angle = 0,
width=1.1\columnwidth]{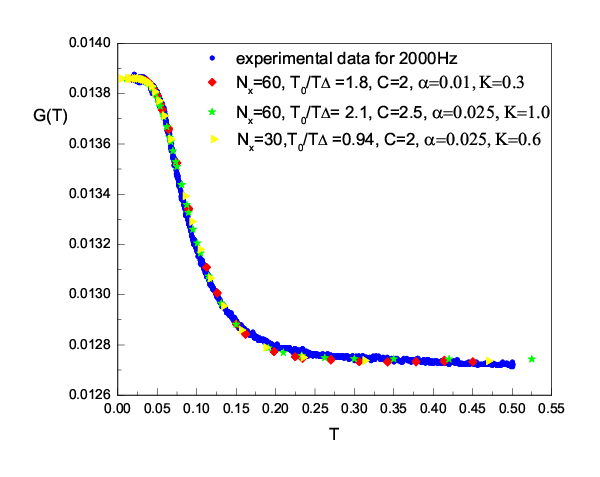}} \caption{(Color online)
Experimental shear modulus $G(T)$ from Ref.\cite{Beamish} and its fits by Eq.(\ref{fit}) for three different sets of parameters, $A=0.0888$.
Error bars are smaller than symbols sizes.
} \label{modulus}
\end{figure}

{\it Finite $T$ behavior}. At $T=0$ there is a gap $\Delta$ in the spectrum of normal excitations. Thus, $T$-dependence $G(T)$ of $G$ becomes significant only at $T\geq \Delta$. As $T$ increases further, $\kappa$ becomes finite and eventually reaches the asymptotic value $\kappa_\infty\equiv \kappa(\infty,L_x)$ defining Eq.(\ref{Gm}). Specific shape of $\kappa(T,L_x)$ has been obtained from the MC simulations of the action (\ref{S_J}) at various $T=T_0/(\Delta_t N_t)$ for fixed $N_x$ ($\Delta_t$ varied as $0.5-1.5$ and it has been checked that the result does not depend on the time discretization). As it turns out, $\kappa$ can be fit by some master curve as well: $\kappa/\kappa_\infty=F(T/T_\Delta)$, with $F(x)$ being some universal function;  dependence on the parameters $C,K,\alpha,N_x,$ is hidden in $T_\Delta$. Thus, for any $T$ Eq.(\ref{Gdual}) becomes,Fig.\ref{modulus}, 
\begin{eqnarray}
G=\frac{G_0}{1+ AF(T/T_\Delta)}, \, A= \left(\frac{G_0}{G_\infty}-1\right).
\label{fit}
\end{eqnarray}

{\it Core superfluidity and shear modulus stiffening}. One of the most striking results of Ref.\cite{Beamish} is the similarity of the temperature dependencies of  $\rho_s(T)$ and $G(T)$. Here we demonstrate that such feature can naturally arise if the core SF is strongly suppressed in the presence of the geometrical kinks formed on segments of the dislocation network at finite $T$. A typical $T$-scale where $\kappa$ becomes finite is given by $\Delta$ so that it is natural to expect that SF is suppressed at $T=\tilde{T}_c \approx \Delta$ rather than at some intrinsic $T_c$ for smooth dislocation, provided $\Delta <<T_c$. Phenomenologically, this implies the term $b' \, \kappa(T)|\psi|^2 $, where $b'>0$ is some coefficient in Landau free energy $H_L$, with $\psi$ standing for an effective 3D SF order parameter. 
The SF part can be written as $-b_0(T) |\psi|^2 + B|\psi|^4/2$, with some constant $B>0$ and function $b_0(T)<0$ for $T>T_c$ and $b_0(T)>0$ for $T<T_c$. If $\tilde{T}_c<<T_c$, one can set $T=0$ in $b_0(T)$ and obtain
\begin{eqnarray}
H_{L}= b_0(0)\left(  \frac{\kappa(T)}{\kappa(\tilde{T}_c)} -1 \right) |\psi|^2 + \frac{B}{2}|\psi|^4,
\label{Hint}
\end{eqnarray}
where $b'$ has been eliminated from the condition that the actual transition occurs at $\tilde{T}_c$.
Thus, using $\rho_s(0)=|\psi(0)|^2 = b_0(0)/B$ (since $\kappa(0)=0$), one finds 
\begin{eqnarray}
\frac{\rho_s(T)}{\rho_s(0)}= 1- \frac{\kappa(T)}{\kappa(\tilde{T}_c)},\,\, 
\label{rhos}
\end{eqnarray}
for $T<\tilde{T}_c$ and $\rho_s(T)/\rho_s(0)=0$ otherwise. 
 We define relative variation of the modulus (\ref{Gdual}): 
$g(T)= (G(T) - G(\tilde{T}_c))/(G(0) - G(\tilde{T}_c))$ and find from Eq.(\ref{Gdual})
\begin{eqnarray}
g(T)= \frac{1- \frac{\kappa(T)}{\kappa(\tilde{T}_c)}}{1+\gamma \kappa(T)} \approx 1- \frac{\kappa(T)}{\kappa(\tilde{T}_c)} 
\label{g}
\end{eqnarray}
within 10\% of accuracy (since $\gamma \kappa(T) \leq 0.1$).
So, Eqs.(\ref{rhos},\ref{g}) imply $\rho_s(T)/\rho_s(0)\approx g(T)$.

Summarizing, we have shown that a single dislocation must be smooth at $T=0$. Crossover to finite temperatures leads to classical roughening and, accordingly, to {\it intrinsic} softening of the shear modulus. Our model contrasts with the model \cite{Beamish}, where the central role in the softening effect is endowed to the \hee3 impurities boiling off from the dislocation cores and, therefore, eliminating pinning centers.  
Within a minimal model, where superfluidity along dislocation core is suppressed by geometrical kinks, it is possible to understand the similarity between SF and dynamical responses \cite{KC,Beamish} of solid \he4. 

Present study focuses on a simplified model where collective effects of the dislocation network are ignored and, therefore, plenty of unanswered questions remains: What is the role of finite density of dislocations, in general, and, specifically, can there be a screening of the long-range interactions for certain type of deformations? Can a 1D crossover to classically rough state at finite $T$ become an actual phase transition in the 3D network? What is the behavior of slanted dislocations? Our model does not address the issue of \hee3 impurities. They provide non-periodic trapping potential for dislocations and, therefore, their role in modifying $G(T)$ requires a separate analysis.

Authors are grateful to John Beamish for providing experimental data and stimulating conversations.
We also acknowledge useful discussions with  Vadim Cheianov, Sebastien Balibar, Nikolay Prokof'ev and Boris Svistunov. One of us (A.B.K.) is thankfull to ICTP, Trieste, for hospitality during Workshop "Supersolid 2008". 
This work was supported by the National Science Foundation
under Grant No. PHY-0653135, CUNY PSC grant and Collaborative grant 80209-0914.
Simulations were performed on  CSI supercomputers -- Athena and Typhon.

\end{document}